\documentclass[aps,prd,twocolumn,showpacs,nofootinbib]{revtex4-1}

\usepackage{graphicx,color}
\usepackage[colorlinks=true, linkcolor=blue, citecolor=blue, urlcolor=blue]{hyperref}

\begin{document}

\title{A novel demonstration of the renormalization group invariance of the fixed-order predictions using the principle of maximum conformality and the $C$-scheme coupling}

\author{Xing-Gang Wu}
\email{wuxg@cqu.edu.cn}

\author{Jian-Ming Shen}
\email{cqusjm@cqu.edu.cn}

\author{Bo-Lun Du}
\email{dblcqu@cqu.edu.cn}

\affiliation{Department of Physics, Chongqing University, Chongqing 401331, P.R. China}

\author{Stanley J. Brodsky}
\email{sjbth@slac.stanford.edu}
\affiliation{SLAC National Accelerator Laboratory, Stanford University, Stanford, California 94039, USA}

\date{\today}

\begin{abstract}

As a basic requirement of the renormalization group invariance, any physical observable must be independent of the choice of both the renormalization scheme and the initial renormalization scale. In this paper, we show that by using the newly suggested $C$-scheme coupling, one can obtain a demonstration that the {\it Principle of Maximum Conformality}  prediction is scheme-independent to all-orders for any renormalization schemes, thus satisfying all of the conditions of the renormalization group invariance. We illustrate these features for the non-singlet Adler function and for $\tau $ decay to $\nu +$ hadrons at the four-loop level.

\begin{description}

\item[PACS numbers] 12.38.Aw, 12.38.Bx, 11.10.Gh, 11.10.Hi

\end{description}

\end{abstract}

\maketitle


\section{Introduction}
\label{sec:intro}

It is well-known that the fixed-order perturbative predictions for observables in Quantum Chromodynamics (QCD) using conventional methods suffer from an uncertainty in fixing the renormalization scale. It is assumed that at sufficiently high order, one will eventually achieve reliable predictions and minimal dependence on the guessed renormalization scale for global quantities such as a total cross-section. A small scale-dependence for the global quantity is generally caused by accidental cancelations among different orders; the scale uncertainty for contributions at each order could still be very large. It is known that the pQCD series will suffer from divergent {\it renormalon} contributions~\cite{Beneke:1998ui, Gardi:2001wg}. Thus even if a perturbative QCD prediction based on a guessed scale agrees with measurements, one cannot be certain that it is a reliable, accurate representation of the theory.

A valid prediction for any physical observable must be independent of the choice of both the initial renormalization scale and the renormalization scheme; this is the central property of {\it renormalization group invariance} (RGI)~\cite{Callan:1970yg, Symanzik:1970rt, Stueckelberg:1953dz, peter2, Peterman:1978tb}. Thus a primary goal for testing pQCD reliably is how to set the renormalization scale such that one obtains accurate fixed-order predictions with maximum precision while satisfying the principle of RGI~\cite{Wu:2013ei}.

The {\it Principle of Maximum Conformality} (PMC)~\cite{Brodsky:2011ig, Brodsky:2011ta, Brodsky:2012sz, Brodsky:2012rj, Mojaza:2012mf, Brodsky:2013vpa} determines the value of the renormalization scale of the QCD running coupling $\alpha_s$ based on the properties of renormalization group equation (RGE). When one applies the PMC, all nonconformal terms are systematically eliminated at each finite order. The PMC satisfies the self-consistency conditions of the renormalization group, such as reflectivity, symmetry and transitivity~\cite{Brodsky:2012ms}. The PMC scales are achieved by applying the RGE recursively. Specific values for the PMC scales are computed as a perturbative expansion, so they have small uncertainties which vary order-by-order. Since the running coupling sums all of the $\{\beta_i\}$-terms, the divergent renormalon terms disappear in the PMC prediction, leading to a more convergent pQCD series. The PMC provides the underlying principle for the Brodsky-Lepage-Mackenzie (BLM) method~\cite{Brodsky:1982gc}, and it reduces to the standard Gell-Mann and Low (GM-L) scale-setting procedure~\cite{GellMann:1954fq} in the $N_c \to 0$ QED Abelian limit~\cite{Brodsky:1997jk}.

To make the PMC scale-setting procedures simpler and more easy to be automatized, a single-scale approach (PMC-s) has been suggested in Ref.\cite{Shen:2017pdu}. This method replaces the individual PMC scales at each order by a single scale in the sense of a mean value theorem; moreover, its predictions are explicitly independent of the choice of initial renormalization scale. Since the coefficients obtained using the PMC-s are identical to those of a conformal theory, one can derive all-orders ``commensurate scale relations" among physical observables~\cite{Brodsky:1994eh, Shen:2016dnq}.

A novel {\it $C$-scheme} coupling has been suggested in the literature, whose scheme-and-scale running behaviors are both governed by a single scheme-independent RGE~\cite{Boito:2016pwf}. In the paper, we shall give the analytic solution for its RGE. Furthermore, by using the $C$-scheme coupling, we shall show that a strict demonstration on the scheme-independence of PMC prediction to all-orders for any renormalization schemes can be achieved. Thus, by combining the $C$-scheme coupling with the PMC-s approach, the resulting predictions become completely independent of the choice of the renormalization scheme and the initial renormalization scale, satisfying all of the conditions of RGI.

The remaining parts of the paper are organized as follows. In Sec.II, we define the $C$-scheme coupling, deduce its RGE and give its solution. We then give the PMC prediction for the $C$-scheme coupling and demonstrate how scheme-and-scale independent predictions can be achieved. In Sec.III, we present numerical results for two examples, the non-singlet Adler function and $\tau $ decays to $\nu +$ hadrons, up to four-loop level. Sec.IV is reserved for a summary.

\section{The PMC predictions under the $C$-scheme coupling}
\label{sec:scalerun}

\subsection{The $C$-scheme coupling and its scheme-and-scale running behavior}

The QCD coupling $\alpha_s(\mu)$ satisfies the RGE:
\begin{equation}
\mu^2 \frac{{\rm d}a_\mu}{{\rm d}\mu^2} \,= \, \beta(a_\mu) \,=\, - a_\mu^2 \sum_{i=0}^{\infty}\beta_i a_\mu^i.
\label{eq:betafun}
\end{equation}
where $a_{\mu}= \alpha_s(\mu)/\pi$, where $\mu$ is the renormalization scale, throughout the paper. Various terms in $\beta_0$, $\beta_1$, $\cdots$, correspond to the one-loop, two-loop, $\cdots$, contributions to the RGE, respectively. The first two terms $\beta_0=(11-{2\over 3} n_f)/4$ and $\beta_1=(102-{38 \over 3}n_f)/4^2$, where $n_f$ is the number of active quarks, are universal in mass-independent schemes; the remaining $\{\beta_i\}$-terms are scheme-dependent. At present, the explicit form for the $\{\beta_i\}$-terms up to five-loop level in the $\overline{\rm MS}$-scheme are available in Refs.~\cite{Gross:1973id, Politzer:1973fx, Caswell:1974gg, Tarasov:1980au, Larin:1993tp, vanRitbergen:1997va, Chetyrkin:2004mf, Czakon:2004bu, Baikov:2016tgj}.

If one integrates the RGE (\ref{eq:betafun}), one obtains
\begin{eqnarray}
&& \ln{\mu_0^2} - \frac{1}{\beta_0 a_{\mu_0}} - \frac{\beta_1}{\beta_0^2}\ln{a_{\mu_0}} - \int_0^{a_{\mu_0}} \frac{{\rm d}a}{\tilde{\beta}(a)} \nonumber\\
&=& \ln \mu^2 - \frac{1}{\beta_0 a_\mu} - \frac{\beta_1}{\beta_0^2}\ln{a_\mu} - \int_0^{a_\mu} \frac{{\rm d}a}{\tilde{\beta}(a)},
\label{eq:scaleInvariant}
\end{eqnarray}
where $\mu_0$ is a reference scale and the $\tilde{\beta}$-function is defined as
\begin{equation}
\frac{1}{\tilde\beta(a)} \,\equiv\, \frac{1}{\beta(a)} + \frac{1}{\beta_0 a^2} - \frac{\beta_1}{\beta_0^2 a}.
\end{equation}
It is useful to define an {\it asymptotic scale} $\Lambda$ by collecting all $\mu_0$-dependent terms on the left-hand-side of Eq.(\ref{eq:scaleInvariant}) into its definition, leading to the evolution of strong coupling $a_\mu$ without reference to a specific choice of $\mu_0$, i.e.
\begin{eqnarray}
\ln\frac{\mu^2}{\Lambda^2} = \frac{1}{\beta_0 a_\mu} + \frac{\beta_1}{\beta_0^2}\ln{a_\mu} + \int_0^{a_\mu} \frac{{\rm d}a}{\tilde{\beta}(a)}.
\label{eq:LambdaQCD}
\end{eqnarray}
The asymptotic scale $\Lambda$ is, by definition, scheme dependent. Eq.(\ref{eq:LambdaQCD}) implies that the conventional $a_\mu$ coupling satisfies the following scheme-dependent scale-running behavior
\begin{eqnarray}\label{eq:general}
\frac{1}{a_\mu} + \frac{\beta_1}{\beta_0}\ln{a_\mu} = \beta_0 \left( \ln\frac{\mu^2}{\Lambda^2} - \int_0^{a_\mu} \frac{{\rm d}a}{\tilde{\beta}(a)} \right).
\end{eqnarray}
As suggested by Ref.\cite{Boito:2016pwf}, one can define a new coupling $\hat{a}_\mu=\hat{\alpha}_s(\mu)/\pi$ in the following way:
\begin{eqnarray}
\frac{1}{\hat a_\mu} + \frac{\beta_1}{\beta_0} \ln\hat a_\mu
&=& \beta_0 \left( \ln\frac{\mu^2}{\Lambda^2} + C \right) \,,
\label{eq:ahat}
\end{eqnarray}
where the phenomenological parameter $C$ is introduced, $C=- \int_0^{a_\mu} {{\rm d}a}/{\tilde{\beta}(a)}$, which incorporates the effects of all scheme-dependent $\{\beta_{i\ge2}\}$-terms. By choosing a specific value for $C$, the running coupling of the $C$-scheme will be equivalent to the coupling of any conventional renormalization scheme.

The solution of Eq.(\ref{eq:ahat}) is much simpler than the one for the conventional RGE that can only be solved via a perturbative way. The solution can be written in terms of the Lambert $W$-function,
\begin{eqnarray}
\hat{a}_\mu = -\frac{\beta_0}{\beta_1 W_{-1}(z)}, \,
z = -\frac{\beta_0}{\beta_1} \exp\left[-\frac{\beta_0^2}{\beta_1}\left( \ln\frac{\mu^2}{\Lambda^2}+C \right)\right],
\end{eqnarray}
where $W_{-1}(z)$ is the solution of $W(z)\exp[W(z)] = z$. The function $W(z)$ is a multi-valued function with an infinite number of branches denoted by $W_n(z)$~\cite{Corless:1996zz}. The correct physical branch can be determined by the requirement that $\hat{a}_\mu$ must be real and positive for a real positive scale $\mu$. Since in practice $n_f\leq6$, we have $z<0$, and the physical branch is $W_{-1}(z)$. One also finds that $W_{-1}(z)$ monotonically decreases within the region of $z\in(-1/e,0)$, with $W_{-1}(z)\in(-\infty,-1)$. The ultraviolet limit corresponds to $z \to 0^{-}$, $W_{-1}(z) \to -\infty$, leading to $\hat{a}_\mu \to 0^{+}$, as required by asymptotic freedom.

Using Eq.(\ref{eq:ahat}), we can obtain a new RGE for the $C$-scheme coupling $\hat{a}_\mu$ which has a much simpler form than the standard RGE (\ref{eq:betafun}):
\begin{equation}
\hat\beta(\hat{a}_\mu) = \mu^2 \frac{\partial \hat{a}_\mu}{\partial \mu^2} = -
\frac{\beta_0 \hat{a}_\mu^2}{ 1 - \frac{\beta_1}{\beta_0} \hat{a}_\mu } = - \beta_0 \hat{a}_\mu^2 \sum_{i=0}^{\infty}\left({\beta_1} /{\beta_0}\right)^{i} \hat{a}_\mu^{i} .
\label{eq:betahat}
\end{equation}
At the same time, from Eq.(\ref{eq:ahat}), one may also observe that
\begin{equation}
\frac{\partial \hat{a}_\mu}{\partial C} = \hat\beta(\hat{a}_\mu).
\label{eq:betahatC}
\end{equation}
Those two equations indicate that
\begin{itemize}
\item Even though the $C$-scheme coupling $\hat{a}_\mu$ itself is implicitly scheme-dependent, its scale-running behavior can be scheme-independent.

\item The $\hat{a}_\mu$ scale-running and scheme-running behaviors separately satisfy the same $\hat\beta$-function.

\item Integrating RGE~(\ref{eq:betahat}) yields a relation of $\hat{a}_\mu$ for any two scales $\mu_1$ and $\mu_2$, i.e.,
 \begin{equation} \label{ahatrun}
  \frac{1}{\hat{a}_{\mu_2}} \,=\, \frac{1}{\hat{a}_{\mu_1}} + {\beta_0}\ln\frac{\mu_2^2}{\mu_1^2}
  - \frac{\beta_1}{\beta_0}\ln\frac{\hat{a}_{\mu_2}}{\hat{a}_{\mu_1}} \,.
 \end{equation}
 Thus if $\hat{a}$ at a scale $\mu_1$ is fixed by a measurement, we can determine its value at any other scale.

\item Given the proper choice of $C$, any coupling constant $a_\mu$ which is defined in any conventional renormalization scheme can be uniquely expressed by a corresponding $C$-scheme coupling $\hat{a}_\mu$.

\end{itemize}

Transforming Eq.(\ref{eq:ahat}) to the following form, we obtain a relation between $\hat{a}_\mu$ and the conventional coupling $a_\mu$,
\begin{displaymath}
\frac{1}{\hat a_\mu} + \frac{\beta_1}{\beta_0} \ln\hat a_\mu = {\beta_0}\,C + \frac{1}{a_\mu} +
\frac{\beta_1}{\beta_0}\ln a_\mu + \beta_0 \!\int_0^{a_\mu}\frac{{\rm d}a}{\tilde\beta(a)},
\end{displaymath}
solving it recursively, we obtain
\begin{eqnarray}
a_\mu &=& \hat{a}_\mu+C\beta_0\hat{a}_\mu^2+ \left(\frac{\beta_2}{\beta_0}- \frac{\beta_1^2}{\beta_0^2}+\beta_0^2 C^2+\beta_1 C\right)\hat{a}_\mu^3 \nonumber\\
& & +\left[\frac{\beta _3}{2 \beta _0}-\frac{\beta_1^3}{2 \beta _0^3}+\left(3 \beta _2-\frac{2 \beta _1^2}{\beta _0}\right) C+\frac{5}{2}\beta_0 \beta_1 C^2 \right. \nonumber\\
&&\left. +\beta_0^3 C^3\right] \hat{a}_\mu^4 + {\cal O}(\hat{a}_\mu^5).
\label{eq:Expandhata}
\end{eqnarray}
This shows that the conventional coupling $a_\mu$ at any scale $\mu$ can be expanded in terms of the $C$-scheme coupling $\hat{a}_\mu$ at the same scale; and vice versa. By choosing a suitable $C$, the new coupling $\hat{a}_\mu$ becomes equivalent to the coupling $a_\mu$ defined for any corresponding conventional scheme; i.e. $a_\mu = \hat{a}_{\mu}|_{C}$. At a different scale $\mu$, a different $C$ needs to be introduced in order to ensure the equivalence of the couplings at the same scale.

\subsection{Scheme-and-scale independent pQCD predictions using PMC scale-setting}
\label{sec:pmc}

The pQCD approximant of an observable up to $n_{\rm th}$-order level can be generally expressed as
\begin{eqnarray}
\rho_{n}(Q) = \sum_{i=1}^{n} r_{i}(\mu/Q) a_{\mu}^{i+p} \label{pQCDexp}
\end{eqnarray}
or
\begin{eqnarray}
\hat\rho_{n}(Q) = \sum_{i=1}^{n} \hat{c}_{i}(\mu/Q) \hat{a}_{\mu}^{i+p}, \label{pQCDexpC}
\end{eqnarray}
where $\mu$ is the renormalization scale and $Q$ is the kinematic scale of the process at which it is measured. Without losing generality, we can set the power of the coupling associated with the tree-level term as $1$, or equivalently $p=0$. The parameters $r_{i}$ and $\hat{c}_{i}$ are the perturbative coefficients for the conventional coupling $a_\mu$ and the corresponding $C$-scheme coupling $\hat{a}_\mu$. Their relations can be obtained by using the relation (\ref{eq:Expandhata}) between the $C$-scheme coupling $\hat{a}_\mu$ and the conventional coupling $a_\mu$,
\begin{eqnarray}
\hat{c}_1 &=& r_1, \label{PMC-s-coeff1}\\
\hat{c}_2 &=& r_2 + \beta_0 r_1 C, \label{PMC-s-coeff2}\\
\hat{c}_3 &=& r_3 + \left(\beta_1 r_1+2\beta_0 r_2\right)C+\beta_0^2 r_1 C^2
+ r_1\left(\frac{\beta_2}{\beta_0}-\frac{\beta_1^2}{\beta_0^2}\right), \label{PMC-s-coeff3} \\
\hat{c}_4 &=& r_4 + \left(3\beta_0 r_3+2\beta_1 r_2+3 \beta_2 r_1-\frac{2\beta_1^2 r_1}{\beta_0}\right)C \nonumber\\
&&
+\left(3\beta_0^2 r_2+\frac{5}{2} \beta_1 \beta_0 r_1\right)C^2+r_1\beta_0^3 C^3 \nonumber \\
&& +r_1\left(\frac{\beta_3}{2\beta_0}-\frac{\beta_1^3}{2\beta_0^3}\right)
 +r_2\left(\frac{2\beta_2}{\beta_0}-\frac{2\beta_1^2}{\beta_0^2}\right).   \label{PMC-s-coeff4}
\end{eqnarray}
The perturbative coefficients $r_{i}$ for the conventional coupling $a_\mu$ can be further expanded as $\{\beta_i\}$-terms, which satisfy the degeneracy relations among different orders~\cite{Brodsky:2013vpa, Mojaza:2012mf, Bi:2015wea}, i.e.,
\begin{eqnarray}
r_1 &=& r_{1,0}, \label{r1-conf-beta}\\
r_2 &=& r_{2,0} + \beta_0 r_{2,1}, \label{r2-conf-beta}\\
r_3 &=& r_{3,0} + \beta_1 r_{2,1} + 2\beta_0 r_{3,1} + \beta _0^2 r_{3,2}, \label{r3-conf-beta}\\
r_4 &=& r_{4,0} + \beta_2 r_{2,1} + 2\beta_1 r_{3,1} + \frac{5}{2} \beta_1 \beta_0 r_{3,2} \nonumber\\
&& +3\beta_0 r_{4,1} + 3 \beta_0^2 r_{4,2} + \beta_0^3 r_{4,3}, \label{r4-conf-beta}
\end{eqnarray}
The non-conformal coefficients $r_{i,j(\geq1)}$ are general functions of $\mu$ and $Q$, which are usually in form of $\ln\mu/Q$. For convenience, we identify the coefficients $r_{i,j(\geq1)}$ as $r_{i,j}=\sum_{k=0}^{j} C_j^k \ln^k(\mu^2/Q^2) \tilde{r}_{i-k,j-k}$, in which $\tilde{r}_{i,j}=r_{i,j}|_{\mu=Q}$ and the combination coefficient $C_j^k={j!}/{k!(j-k)!}$. The conformal coefficients are free from $\mu$-dependence, e.g., $r_{i,0} \equiv \tilde{r}_{i,0}$.

The authors of Refs.\cite{Boito:2016pwf, Jamin:2016ihy} have investigated the possibility of obtaining an ``optimized" prediction for the truncated pQCD series using the $C$-scheme coupling by exploiting its scheme dependence. In their treatment, by fixing $\mu\equiv Q$ and varying $C$ within a possible domain, an optimal $C$-value, and thus an optimal scheme, is determined by requiring the absolute value of the last known term $\hat{c}_n(Q/Q)\hat{a}_Q^n$ to be at its minimum. However, we have noted that the idea of requiring the magnitude of the last known term of the pQCD series to be at its minimum is similar to the postulate of the {\it Principle of Minimum Sensitivity} (PMS)~\cite{Stevenson:1980du, Stevenson:1981vj, Stevenson:1982qw}, in which the optimal scheme is determined by directly requiring all unknown higher-order terms to vanish. Thus this application of optimization to the $C$-scheme coupling approach meets the same problems of PMS, such as it does not satisfy the self-consistency conditions of the renormalization group and etc~\cite{Ma:2014oba, Wu:2014iba}. Moreover, the determined optimal value of $C$ shall be different for a different fixed-order prediction, thus it needs to be redetermined when new perturbative terms are known. Although this approach of using the $C$-scheme coupling could be considered as a practical way to improve pQCD precision, similar to the PMS approach, it cannot be considered as the solution to the conventional scheme-and-scale ambiguities.

In contrast to the PMS, the PMC identifies all the RG-involved scheme-dependent $\{\beta_i\}$-terms in the perturbative series and eliminate them by shifting the scales of the running coupling. After applying the PMC, the coefficients of $\rho_n$ match the corresponding conformal series, and thus the prediction is scheme independent in general. We have demonstrated that the PMC leads to scheme-independent pQCD predictions for any dimensional-like renormalization scheme~\cite{Mojaza:2012mf, Brodsky:2013vpa}.

In the following, we shall demonstrate that one can eliminate all scheme-dependent $C$-terms in a pQCD approximant by applying the PMC-s approach~\cite{Shen:2017pdu}. Since the parameter $C$ identifies any choice of the renormalization scheme, we will achieve a general demonstration of the scheme-independence of the PMC pQCD predictions for any renormalization scheme.

Our demonstration starts from Eq.(\ref{pQCDexpC}), and we shall consider at least the next-to-leading order pQCD correction to the pQCD prediction; i.e. $n\ge2$. Up to four-loop level, the coefficients $\hat{c}_i$ can be related to the coefficients $r_{i,j}$ for conventional running coupling from Eqs.(\ref{PMC-s-coeff1}, \ref{PMC-s-coeff2}, \ref{PMC-s-coeff3}, \ref{PMC-s-coeff4}, \ref{r1-conf-beta}, \ref{r2-conf-beta}, \ref{r3-conf-beta}, \ref{r4-conf-beta}). To avoid complexity for determining the coefficients of each $\{\beta_i\}$-terms, we adopt the PMC-s approach to eliminate all the RG-involved $\{\beta_i\}$-terms.

With the help of Eq.(\ref{eq:betahatC}, \ref{pQCDexpC}), the $C$-dependence of the pQCD approximate $\hat{\rho}_n$ can be expressed as
\begin{equation}
\frac{\partial \hat{\rho}_n}{\partial C} = -\hat{\beta}(\hat{a}_\mu) \frac{\partial \hat{\rho}_n}{\partial \hat{a}_\mu},
\label{eq:Cdependence}
\end{equation}
which shows that when the non-conformal terms associated with the $\hat\beta(\hat{a}_{\mu})$-function have been removed, one can achieve a scheme-independent prediction at any fixed order, i.e. $\hat\beta(\hat{a}_{\mu})\to 0$ indicates ${\partial \hat{\rho}_n}/{\partial C}\to 0$.

Following the standard PMC-s scale-setting procedures, an effective PMC scale $Q_\star$ can be obtained by eliminating all nonconformal terms, which can be expanded as a power series in $\hat{a}_{Q_\star}$ via the following way,
\begin{eqnarray}
\ln\frac{Q_{\star}^2}{Q^2} = \sum^{n-2}_{i= 0} \hat{S}_i \hat{a}^i_{Q_\star},
\label{eq:C-PMC-scale}
\end{eqnarray}
whose first three coefficients are
\begin{eqnarray}
\hat{S}_0 &=& -\frac{\tilde{r}_{2,1}}{\tilde{r}_{1,0}}-C, \nonumber\\
\hat{S}_1 &=& \frac{2\left(\tilde{r}_{2,0}\tilde{r}_{2,1}-\tilde{r}_{1,0}\tilde{r}_{3,1}\right)}{\tilde{r}_{1,0}^2}
+\frac{\tilde{r}_{2,1}^2-\tilde{r}_{1,0} \tilde{r}_{3,2}}{\tilde{r}_{1,0}^2}\beta_0 +\frac{\beta_1^2}{\beta_0^3}-\frac{\beta_2}{\beta_0^2}, \nonumber \\
\hat{S}_2 &=& \frac{3\tilde{r}_{1,0}\left(\tilde{r}_{3,0}\tilde{r}_{2,1} -\tilde{r}_{1,0}\tilde{r}_{4,1}\right)+4\tilde{r}_{2,0} \left(\tilde{r}_{1,0}\tilde{r}_{3,1}-\tilde{r}_{2,0} \tilde{r}_{2,1}\right)}{\tilde{r}_{1,0}^3} \nonumber\\
&&+\frac{3\tilde{r}_{1,0}\tilde{r}_{2,1}\tilde{r}_{3,2} -\tilde{r}_{1,0}^2\tilde{r}_{4,3} -2\tilde{r}_{2,1}^3}{\tilde{r}_{1,0}^3}\beta_0^2 \nonumber\\
&& \hspace{-0.3cm} +\frac{3\tilde{r}_{1,0}\left(2\tilde{r}_{2,1} \tilde{r}_{3,1} -\tilde{r}_{1,0} \tilde{r}_{4,2}\right)+ \tilde{r}_{2,0}\left(2 \tilde{r}_{1,0} \tilde{r}_{3,2}-5 \tilde{r}_{2,1}^2\right)}{\tilde{r}_{1,0}^3}\beta_0 \nonumber\\
&& + \frac{3\left(\tilde{r}_{2,1}^2-\tilde{r}_{1,0}\tilde{r}_{3,2}\right)}{2\tilde{r}_{1,0}^2}\beta_1- \frac{\beta_1^3}{2\beta_0^4}+\frac{\beta_2 \beta_1}{\beta_0^3}-\frac{\beta_3}{2\beta_0^2}. \nonumber
\end{eqnarray}
It is interesting to find that all high-order coefficients $\hat{S}_{i}$ ($i\geq1$) are free of the scheme parameter $C$. By using the definition of $C$-scheme coupling (\ref{eq:ahat}), we obtain
\begin{eqnarray}
\frac{1}{\hat{a}_{Q_\star}}+\frac{\beta_1}{\beta_0} \ln\hat{a}_{Q_\star} &=& \beta_0\left(\ln\frac{Q_\star^2}{\Lambda^2}+C\right)\nonumber\\
&=& \beta_0 \left( \ln\frac{Q^2}{\Lambda^2}-\frac{\tilde{r}_{2,1}}{\tilde{r}_{1,0}} +\sum^{n-2}_{i= 1} \hat{S}_i \hat{a}^i_{Q_\star} \right). \label{Csas2}
\end{eqnarray}
The second equation shows that, even though the effective scale $Q_{\star}$ depends on the choice of $C$, the coupling $\hat{\alpha}_{Q_\star}$ is independent to $C$ at any fixed order.

Thus, after fixing the scale $Q_\star$, we achieve a $C$-scheme independent pQCD series
\begin{eqnarray}
\hat{\rho}_n(Q)|_{\rm PMC} &=& \sum^{n}_{i=1} r_{i,0} \hat{a}_{Q_\star}^i .
\label{eq:PMCprediction}
\end{eqnarray}
The pQCD series depends on the choice of scheme via the coefficients $r_{i,j}$ and the $\{\beta_{i\geq2}\}$-functions. Thus, Eq.(\ref{eq:PMCprediction}) indicates the scheme-independence of the $C$-scheme predictions is equivalent to the scheme-independence of the initial choice of scheme, and vice versa. The demonstration of $C$-scheme independence, as shown by Eq.(\ref{eq:Cdependence}), shows the pQCD predictions are scheme independent for any choice of the initial scheme. By using the above formulas, we obtain pQCD predictions independent of any choice of scheme (represented by any choice of $C$). This demonstrates to any orders the scheme-independent of the PMC predictions -- Given one measurement which sets the value of the coupling at a scale, the resulting PMC predictions are independent of the choice of renormalization scheme.

\section{Phenomenological examples}
\label{sec:PhApp}

In doing the numerical calculations below, we adopt the world average $\alpha_s^{\overline{\rm MS}}(M_Z) =0.1181(11)$~\cite{Olive:2016xmw} as the reference value for fixing the running coupling, which runs down to $\alpha_s^{\overline{\rm MS}}(M_\tau)=0.3159(95)$. $M_Z=91.1876$ GeV and $M_\tau=1.777$ GeV.

\subsection{The non-singlet Adler function}
\label{subsec:Adler}

The non-singlet Adler function \cite{Adler:1974gd} reads
\begin{eqnarray}
D^{\rm ns}(Q^2,\mu) &=& 12\pi^2 \left[\gamma^{\rm ns}(a_\mu) - \beta(a_\mu)\frac{\partial} {\partial{a_\mu}}{\Pi}^{\rm ns}(L,{a_\mu})\right] \nonumber\\
&=& \frac{3}{4}\gamma^{\rm ns}_0 + \bar{D}^{\rm ns}(Q^2,\mu),
\label{eq:nsAdler}
\end{eqnarray}
where $\mu$ is the renormalization scale, $a_\mu=\alpha_s(\mu)/\pi$, and $L=\ln{\mu^2}/{Q^2}$. $\gamma^{\rm ns}(a_\mu) = \sum_{i \ge 0} {\gamma^{\rm ns}_i} {a_\mu^i}/{16\pi^2}$ is the non-singlet part of the photon field anomalous dimension and $\Pi^{\rm ns}(L,a_\mu)=\sum_{i \ge 0} {\Pi_i^{\rm ns}} {a_\mu^i}/{16\pi^2}$ is the non-singlet part of the polarization function for a flavor-singlet vector current. The pQCD series of $\bar{D}^{\rm ns}(Q^2,\mu)$ up to $n_{\rm th}$-loop level can be written as
\begin{eqnarray}
\bar{D}^{\rm ns}_n(Q^2,\mu) = \sum_{i=1}^n r_i(\mu/Q) a_\mu^i.
\end{eqnarray}
The perturbative coefficients $\gamma^{\rm ns}_i$ and $\Pi^{\rm ns}_i$ within the $\overline{\rm MS}$-scheme have been given up to four-loop level~\cite{Baikov:2012zm}, and the coefficients $r_i$ for $\mu=Q$ within the $\overline{\rm MS}$-scheme up to four-loop level can be read from Refs.\cite{Chetyrkin:1996ia, Baikov:2010je}. The coefficients at any other choices of the renormalization scale ($\mu\neq Q$) can be obtained via RGE.

\subsubsection{Predictions using conventional scale-setting}

Taking $Q=M_\tau$, we obtain a four-loop $\overline{\rm MS}$-scheme prediction on $\bar{D}^{\rm ns}$ using conventional scale-setting (Conv.),
\begin{equation}
\bar{D}^{\rm ns}_4(M_\tau^2,\mu=M_\tau)|_{\rm Conv.} = 0.1286 \pm 0.0053 \pm 0.0094,
\label{eq:Adler-MSbar-4loop}
\end{equation}
where the first error is for $\Delta\alpha_s^{\overline{\rm MS}}(M_Z)=\pm0.0011$ and the second error is an estimate of the ``unknown'' high-order contribution, which is conservatively taken as the maximum value of the last known term of the perturbative series within the possible choices of initial scale~\cite{Wu:2014iba}. As for the four-loop prediction, we take the maximum value of $|r_4(\mu/M_{\tau}) a_\mu^4|$ with $\mu\in[M_{\tau},4M_{\tau}]$ as the estimated ``unknown'' high-order contribution.

The unknown fifth-order coefficient has been estimated by several groups, e.g. $r_5\simeq 283$~\cite{Beneke:2008ad} or $r_5\simeq 275$~\cite{Baikov:2008jh}. If using $r_5\simeq 283$, Eq.(\ref{eq:Adler-MSbar-4loop}) changes to
\begin{equation}
\bar{D}^{\rm ns}_5(M_\tau^2,\mu=M_\tau)|_{\rm Conv.} = 0.1315 \pm 0.0057 \pm 0.0065.
\label{eq:Adler-MSbar-5loop}
\end{equation}

In addition to the scale dependence, the predictions using conventional scale setting is also scheme dependent at any fixed order. We adopt the $C$-scheme coupling to illustrate this dependence, and we can rewrite $\bar{D}^{\rm ns}_n(Q^2,\mu)$ in terms of the $C$-scheme coupling $\hat{a}_\mu$ as
\begin{eqnarray}
\bar{D}_n^{\rm ns}(Q^2,C) = \sum_{i=1}^n \hat{c}_i(\mu/Q) \hat{a}_\mu^i,
\end{eqnarray}
where the coefficients $\hat{c}_i(\mu/Q)$ can be derived by using Eqs.(\ref{PMC-s-coeff1}, \ref{PMC-s-coeff2}, \ref{PMC-s-coeff3}, \ref{PMC-s-coeff4}).

\begin{figure}[htb]
\includegraphics[width=0.45\textwidth]{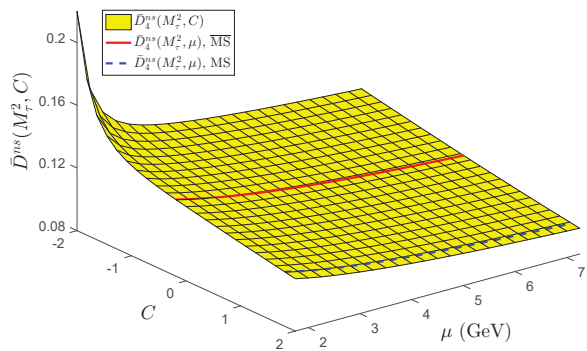}
\caption{The four-loop prediction on $\bar{D}^{\rm ns}(M_\tau^2,C)$ for the Adler function using conventional scale setting, which is shown by a light shaded band. The solid line is for the $\overline{\rm MS}$-scheme, and the dashed line is for the MS-scheme.}
\label{fig:Adler4l3D}
\end{figure}

A graphical representation of $\bar{D}^{\rm ns}(M_\tau^2,C)$ as a function of parameters $C$ and $\mu$ is given in Fig.~\ref{fig:Adler4l3D}, in which we have chosen $C\in[-2,+2]$~\footnote{The relation between the $C$-scheme coupling $\hat{a}_{M_\tau}$ and the $\overline{\rm MS}$-scheme coupling $a_{M_\tau}$ ceases to be perturbative and breaks down below $C\sim -2$. Thus in our discussions we shall adopt $C\ge -2$.} and $\mu \in \left[M_\tau, 4M_\tau\right]$. The shaded band shows the four-loop prediction on $\bar{D}^{\rm ns}(M_\tau^2,C)$, which has a large scheme-and-scale dependence. Using a proper $C$, the prediction using the $C$-scheme coupling $\hat{a}_\mu$ is equivalent to predictions using some of the familiar schemes; e.g. the solid line in Fig.\ref{fig:Adler4l3D} is for the $\overline{\rm MS}$-scheme and the dashed line is for the MS-scheme. To ensure the equivalence, the value of $C$ should be changed for different scales. By taking $C=-0.188$ one obtains the conventional $\overline{\rm MS}$ prediction for $\mu=M_\tau$; alternatively it can taken as $C=-0.004$ for $\mu=4M_\tau$.

Requiring the estimated ``unknown'' high-order contribution, $|\hat{c}_n(\mu/M_\tau) \hat{a}_{\mu}^n|_{\rm MAX}$, to be at its minimum, we can obtain an optimal $C$-scheme for $\bar{D}_n^{\rm ns}(Q^2,C)$. At the four-loop level with $n=4$, the optimal $C$-value is $-0.972$, leading to
\begin{displaymath}
\bar{D}^{\rm ns}_4(M_\tau^2,C=-0.972)|_{\rm Conv.} = 0.1365 \pm 0.0069 \pm 0.0083,
\end{displaymath}
where the central value is for $\mu=M_\tau$, the first error is for $\Delta\alpha_s^{\overline{\rm MS}}(M_Z)=\pm0.0011$ and the second error is an estimate of the ``unknown'' high-order contribution. As for a five-loop prediction, if using $r_5\simeq 283$, the optimal $C$-value changes to $-1.129$, and we obtain
\begin{displaymath}
\bar{D}^{\rm ns}_5(M_\tau^2,C=-1.129)|_{\rm Conv.} = 0.1338 \pm 0.0062 \pm 0.0054.
\end{displaymath}

\subsubsection{Predictions using PMC scale-setting}

Following the standard PMC-s procedures, by resumming all the RG-involved non-conformal $\{\beta_i\}$-terms into the running coupling, we obtain
\begin{equation}
\bar{D}^{\rm ns}_n(Q^2,C)|_{\rm PMC} = \frac{3}{4} \sum_{i=1}^n \gamma^{\rm ns}_i \hat{a}_{Q_\star}^{i}.
\end{equation}
Using the known four-loop pQCD prediction $\bar{D}^{\rm ns}_4$, the PMC scale $Q_\star$ can be determined up to next-to-next-to-leading log (N$^2$LL) accuracy:
\begin{equation}
\ln\frac{Q_{\star}^2}{Q^2} = -C+0.2249 -3.1382\hat{a}_{Q_\star} -13.3954\hat{a}_{Q_\star}^2,
\label{eq:AdlerPMCscale}
\end{equation}
in which the value of the $C$-scheme coupling $\hat{a}_{Q_\star}$ can be determined by Eq.(\ref{Csas2}).

\begin{figure}[htb]
\centering
\includegraphics[width=0.45\textwidth]{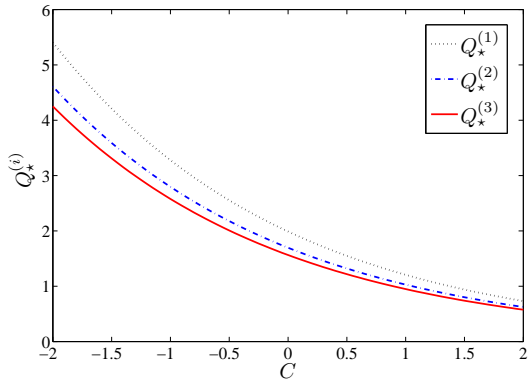}
\caption{The PMC scale $Q^{(i)}_\star$ of $\bar{D}_4^{\rm ns}(M_\tau^2,C)$ as a function of parameter $C$, where $i$ indicates the scale is at the N$^{(i-1)}$LL accuracy. }
\label{fig:AdlerS}
\end{figure}

The PMC scale $Q_\star$ is independent of the choice of the initial scale, which depends on $C$. We present $Q_\star$ as a function of $C$ in Fig.\ref{fig:AdlerS}, in which $Q^{(1,2,3)}_\star$ are at the LL, NLL and ${\rm N^{2}LL}$ level, respectively. Fig.\ref{fig:AdlerS} shows $Q^{(1,2,3)}_\star$ decreases with the increment of $C$, and $Q^{(1)}_\star > Q^{(2)}_\star > Q^{(3)}_\star$. The optimal scale $Q_\star$ is of perturbative nature: when more loop terms are included, it becomes more accurate. Eq.(\ref{Csas2}) shows that the $C$-scheme coupling is independent to the choice of $C$. By taking $Q=M_\tau$, we obtain $\hat{a}_{Q_\star} \equiv 0.1056(41)$ for any choice of $C$, where the errors are from $\Delta\alpha_s^{\overline{\rm MS}}(M_Z)=\pm0.0011$. We then obtain the scheme-independent PMC prediction on $\bar{D}_4^{\rm ns}$,
\begin{eqnarray}
\bar{D}_4^{\rm ns}(M_\tau^2,C)|_{\rm PMC} = 0.1345 \pm 0.0066 \pm 0.0008,
\label{eq:Adler-PMC}
\end{eqnarray}
where the first error is for $\Delta\alpha_s^{\overline{\rm MS}}(M_Z)=\pm0.0011$ and the second error is an estimate of the ``unknown'' high-order contribution.

\begin{figure}[htb]
\includegraphics[width=0.45\textwidth]{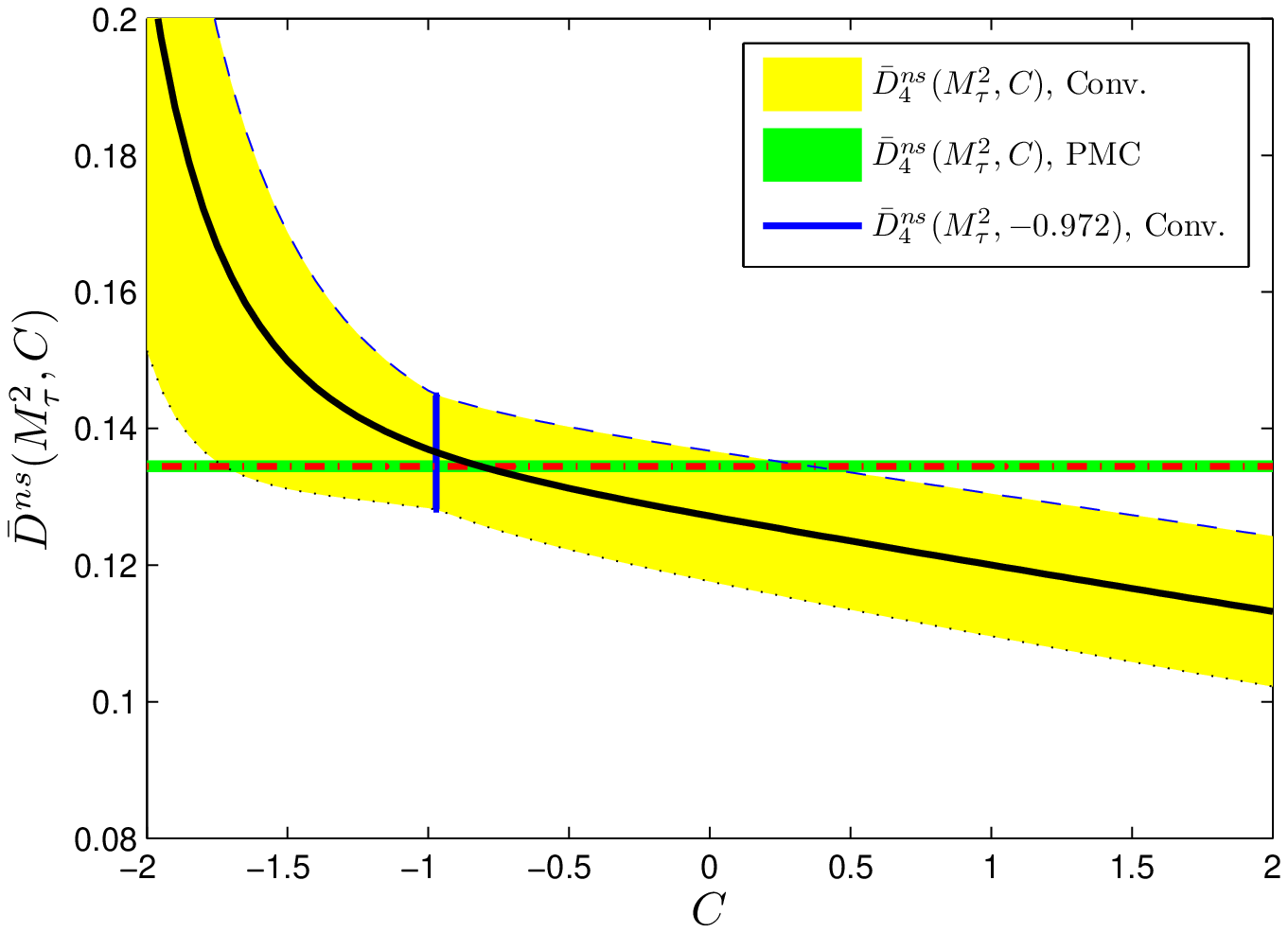}
\includegraphics[width=0.45\textwidth]{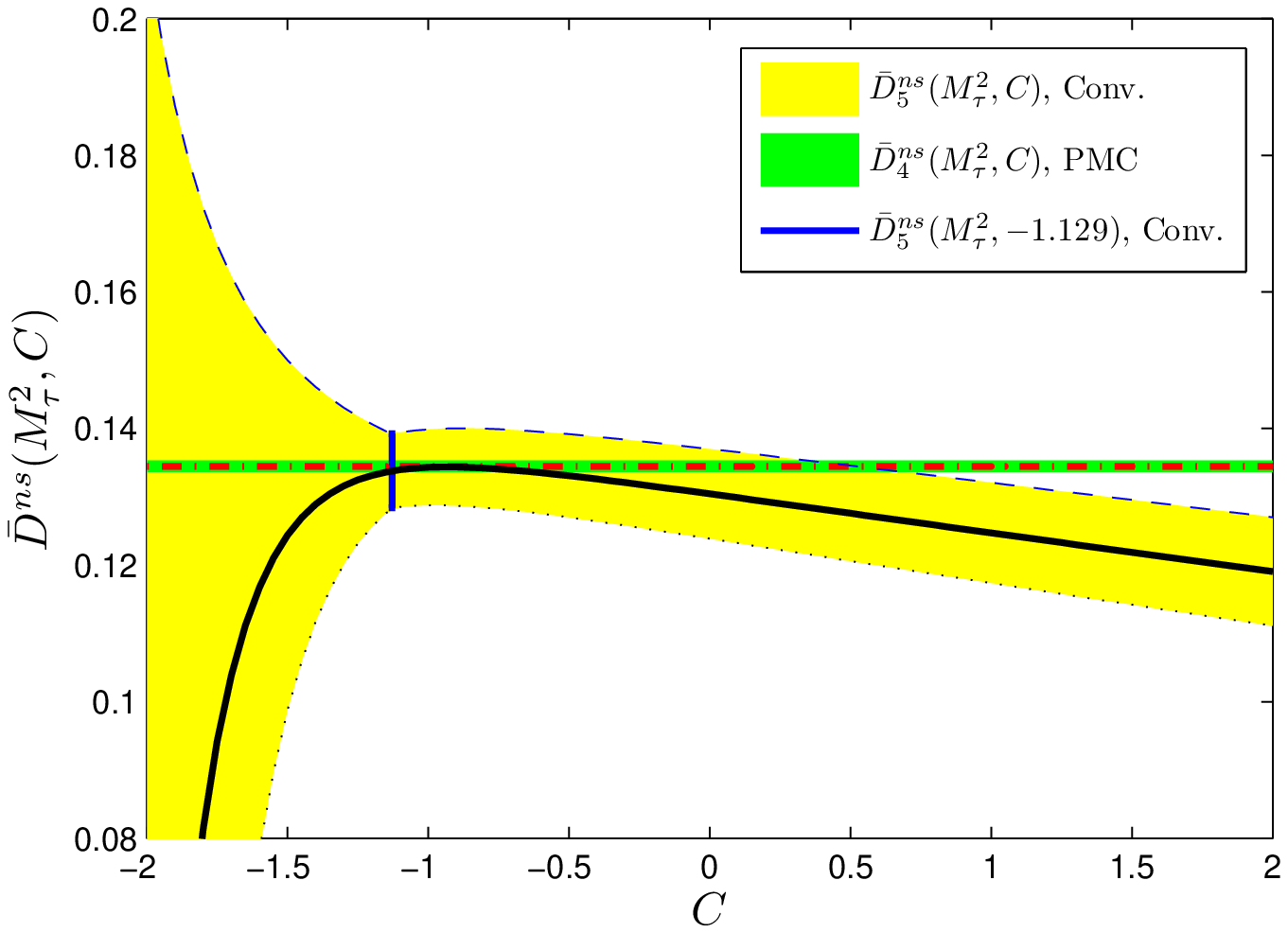}
\caption{$\bar{D}^{\rm ns}(M_\tau^2,C)$ for the Adler function as a function of the parameter $C$. The solid line is the prediction using conventional scale setting, the lighter-shaded band is the uncertainty for a four-loop prediction $\Delta = \pm|\hat{c}_4(\mu/M_\tau) \hat{a}_{\mu}^4|_{\rm MAX}$ (Upper) and for an approximate five-loop prediction $\Delta = \pm|\hat{c}_5(\mu/M_\tau)\hat{a}_{\mu}^5|_{\rm MAX}$ (Lower), where MAX is the maximum value for $\mu\in[M_\tau,4M_{\tau}]$. When $C=-0.972$ (Upper) and $C=-1.129$ (Lower), the error bar as shown by a vertical solid line is the minimum. The dash-dot line represents the four-loop PMC prediction, and the darker shaded band is for $\Delta = \pm |r_{4,0}\hat{a}^4_{Q_\star}|$ . }
\label{fig:Adler4l2Dimproved}
\end{figure}

In order to compare the scheme dependence before and after applying the PMC, we present the predictions for $\bar{D}^{\rm ns}(M_\tau^2,C)$ for $\mu=M_\tau$ in Fig.~\ref{fig:Adler4l2Dimproved}. At four-loop level, Fig.~\ref{fig:Adler4l2Dimproved} shows that the scheme-dependence of $\bar{D}^{\rm ns}_4(M_\tau^2,C)|_{\rm Conv.}$ is rather large, which decreases with increasing values for $C$; for larger $C$ values, the error band becomes slightly larger. When $C=-0.972$, the error bar is the minimum, corresponding to the optimal scheme. By using the approximate five-loop term $r_5\simeq 283$, we also give the results for the approximate five-loop prediction. Fig.~\ref{fig:Adler4l2Dimproved} shows a smaller error bar is achieved with a five-loop term, $\bar{D}^{\rm ns}_5(M_\tau^2,C)|_{\rm Conv.}$ first increases and then decreases with the increment of $C$, and the optimal scheme is slightly shifted to $C=-1.129$. It is interesting to find that the predicted $\bar{D}^{\rm ns}(M_\tau^2,C)|_{\rm Conv.}$ with the optimal choice of $C$ are consistent with the PMC predictions within errors. This indicates that the differences caused by various scale-setting approaches could be smeared by including more and more loop terms and be consistent with the scheme-independent predictions. The flat dash-dot line in Fig.~\ref{fig:Adler4l2Dimproved} shows that the conventional scheme dependence can be eliminated by applying the PMC. Due to the much faster pQCD convergence after applying the PMC, and due to the elimination of the scale dependence, the PMC suggests that the unknown high-order contribution could be quite small in comparison to the present four-loop prediction.

\begin{table*}[htb]
\centering
\caption{The value of each loop-term, LO, NLO, N$^2$LO, or N$^3$LO, for the four-loop prediction $\bar{D}^{\rm ns}_4$ using conventional (Conv.) and PMC scale-settings, respectively. $\mu=Q=M_\tau$. The results for the $\overline{\rm MS}$-scheme and the optimal $C$-scheme with $C=-0.972$ are presented. The PMC prediction is unchanged for any $C$-scheme. }
\begin{tabular}{ c c c c c c c c}
\hline
~~~~   & ~LO~ & ~NLO~ & ~N$^2$LO~ & ~N$^3$LO~ & ~Total~ \\
\hline
~Conv., $\overline{\rm MS}$-scheme~  & 0.1006 & 0.0166 & 0.0064 & 0.0050 & 0.1286  \\
~Conv., optimal $C$-scheme~ & 0.1347 & $-0.0099$ & 0.0034 & 0.0083 & 0.1365  \\
\hline
~PMC, any $C$-scheme~  & 0.1056 & 0.0240 & 0.0041 & 0.0008 & 0.1345  \\
\hline
\end{tabular}
\label{tab:AdlerOrderNew}
\end{table*}

We present the value of each loop-term, LO, NLO, N$^2$LO, or N$^3$LO, for the four-loop prediction $\bar{D}^{\rm ns}_4$ using conventional and PMC scale-settings in Table~\ref{tab:AdlerOrderNew}. The pQCD convergence for the conventional $\overline{\rm MS}$-scheme is moderate. The pQCD convergence for the optimal $C$-scheme ($C=-0.972$) does not suffer from the usual $\alpha_s$-suppression, the relativity of the related high-loop terms show, $|\bar{D}^{\rm ns, LO}_{4}| \gg |\bar{D}^{\rm ns, NLO}_{4}| \sim |\bar{D}^{\rm ns, N^2LO}_{4}| \sim |\bar{D}^{\rm ns, N^3LO}_{4}|$. On the other hand, by applying the PMC, a much better pQCD convergence is achieved due to the elimination of divergent renormalon-like terms.

\subsection{$\tau$ decays to $\nu+$ hadrons}
\label{subsec:taudecay}

The ratio of the $\tau$ total hadronic branching fraction to its lepton branching fraction can be parameterized as,
\begin{eqnarray}
R_{\tau} &=& \frac{\Gamma(\tau\rightarrow{\rm hadrons}+\nu_\tau)}{\Gamma(\tau\rightarrow l+\bar\nu_l+\nu_\tau)} \nonumber\\
&=& 3 S_{\rm EW} \left( |V_{ud}|^2 + |V_{us}|^2 \right)(1+\delta^{(0)} + \cdots),
\end{eqnarray}
where $S_{\rm EW}$ is an electroweak correction, $|V_{ud}|$ as well as $|V_{us}|$ are CKM matrix elements. The pQCD correction is encoded in $\delta^{(0)}$ and the ellipsis indicate further small subleading corrections. The pQCD correction up to $n_{\rm th}$-order level reads,
\begin{eqnarray}
\delta^{(0)}_n(M_\tau^2,\mu) = \sum_{i=1}^n c_i(\mu/M_\tau) a_\mu^i,
\end{eqnarray}
where the coefficients $c_i$ for the $\overline{\rm MS}$-scheme up to four-loop level can be found in Refs.\cite{Baikov:2008jh, Beneke:2008ad}.

\subsubsection{Predictions using conventional scale-setting}

Taking $\mu=M_\tau$, we obtain a four-loop $\overline{\rm MS}$-scheme prediction on $\delta^{(0)}$ using conventional scale-setting,
\begin{equation}
\delta^{(0)}_4(M_\tau^2,\mu=M_\tau)|_{\rm Conv.} = 0.1930 \pm 0.0104 \pm 0.0199,
\label{eq:RtauMSbar-4loop}
\end{equation}
where the first error is for $\Delta\alpha_s^{\overline{\rm MS}}(M_Z)=\pm0.0011$ and the second error is an estimate of the ``unknown'' high-order contribution. If we use the predicted five-loop term $r_5\simeq 283$~\cite{Beneke:2008ad} , we obtain
\begin{equation}
\delta^{(0)}_5(M_\tau^2,\mu=M_\tau)|_{\rm Conv.} = 0.1990 \pm 0.0113 \pm 0.0151.
\label{eq:RtauMSbar-5loop}
\end{equation}

\begin{figure}[htb]
\includegraphics[width=0.45\textwidth]{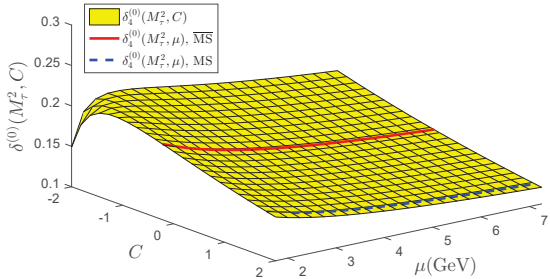}
\caption{The four-loop prediction on $\delta^{(0)}(M_\tau^2,C)$ using conventional scale setting, which is shown by a light shaded band. The solid line is for the $\overline{\rm MS}$-scheme, and the dashed line is for the MS-scheme. }
\label{fig:Rtau4l3D}
\end{figure}

The predictions using conventional scale setting are scheme dependent. By using the relation (\ref{eq:Expandhata}), we rewrite $\delta^{(0)}_n(M_\tau^2,\mu)$ in terms of the $C$-scheme coupling $\hat{a}_\mu$ as
\begin{eqnarray}
\delta^{(0)}_n(M_\tau^2,C) = \sum_{i=1}^n \hat{c}_i(\mu/M_\tau) \hat{a}_\mu^i .
\end{eqnarray}
A graphical representation of $\delta^{(0)}(M_\tau^2,C)$ as a function of parameters $C$ and $\mu$ is given in Fig.~\ref{fig:Rtau4l3D}, in which we have chosen $C\in[-2,+2]$ and $\mu \in \left[M_\tau, 4M_\tau\right]$. The shaded band shows the four-loop prediction on $\delta^{(0)}(M_\tau^2,C)$, the solid line is $\overline{\rm MS}$-scheme prediction and the dashed line is the MS-scheme one.

Requiring the approximate ``unknown'' high-order contribution, $|\hat{c}_n(\mu/M_\tau) \hat{a}_{\mu}^n|_{\rm MAX}$, minimal, we obtain an optimal $C$-scheme for $\delta^{(0)}(M_\tau^2,C)$ at the $n_{\rm th}$-order level. Using the four-loop prediction with $n=4$, the optimal $C$-value is $-1.638$, which leads to
\begin{displaymath}
\delta^{(0)}_4(M_\tau^2,C=-1.638)|_{\rm Conv.} = 0.1979 \pm 0.0099 \pm 0.0186,
\end{displaymath}
where the first error is for $\Delta\alpha_s^{\overline{\rm MS}}(M_Z)=\pm0.0011$, and the second error is for $\pm|\hat{c}_4(\mu/M_\tau) \hat{a}_{\mu}^4|_{\rm MAX}$. If using the five-loop term $r_5 \simeq 283$, the optimal $C$-value changes to $-1.813$, and we obtain
\begin{displaymath}
\delta^{(0)}_5(M_\tau^2,C=-1.813)|_{\rm Conv.} = 0.1968 \pm 0.0095 \pm 0.0118.
\end{displaymath}

\subsubsection{Predictions for $\tau $ decay using PMC scale-setting}

Up to four-loop level, the known conformal and non-conformal coefficients for conventional coupling can be read from Ref.\cite{Brodsky:2013vpa}. By resumming all the RG-involved non-conformal $\{\beta_i\}$-terms into the running coupling, we obtain the PMC prediction for $\delta^{(0)}_n(M_\tau^2,C)$, i.e.
\begin{eqnarray}
\delta^{(0)}_n(M_\tau^2,C)|_{\rm PMC} = \frac{3}{4} \sum_{i=1}^n \gamma^{\rm ns}_i \hat{a}_{Q_\star}^i,
\end{eqnarray}
Using the known four-loop pQCD prediction $\delta^{(0)}_4(M_\tau^2,C)$, the PMC scale $Q_\star$ can be determined up to next-to-next-to-leading log (N$^2$LL) accuracy:
\begin{eqnarray}
\ln\frac{Q_{\star}^2}{M_\tau^2} = -C-1.3584+1.6234\hat{a}_{Q_\star}-1.1385\hat{a}_{Q_\star}^2,
\label{eq:RtauPMCscale}
\end{eqnarray}
where the optimal $C$-scheme coupling $\hat{a}_{Q_\star}$ is determined by Eq.(\ref{Csas2}).

\begin{figure}[htb]
\includegraphics[width=0.45\textwidth]{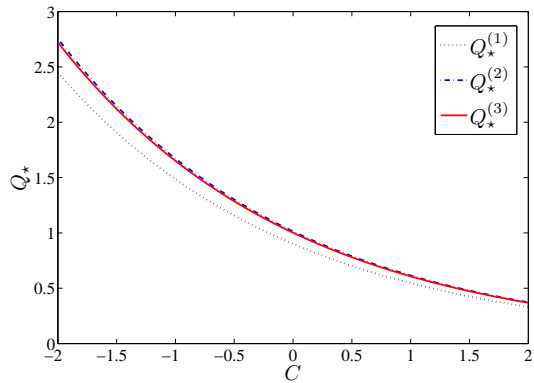}
\caption{The PMC scale $Q^{(i)}_\star$ of $\delta^{(0)}_4(M_\tau^2,C)$ as a function of parameter $C$, where $i$ indicates the scale is at the N$^{(i-1)}$LL accuracy. }
\label{fig:Rtau-PMC-scale}
\end{figure}

We present the scheme-dependent $Q_\star$ as a function of $C$ in Fig.\ref{fig:Rtau-PMC-scale}, in which $Q^{(1,2,3)}_\star$ are at the LL, NLL and ${\rm N^{2}LL}$ level, respectively. To compare with the case of $\bar{D}^{\rm ns}(Q^2)$, the perturbative series for $\ln{Q^2_\star}/{M_\tau^2}$ oscillates, leading to $Q^{(1)}_\star < Q^{(2)}_\star$ and $Q^{(2)}_\star > Q^{(3)}_\star$; similar to the case of $\bar{D}^{\rm ns}(Q^2)$, $Q^{(1,2,3)}_\star$ decreases with increasing $C$. Eq.(\ref{Csas2}) indicates the $C$-scheme coupling $\hat{a}_{Q_\star}$ is free of the parameter $C$, and for the four-loop level we obtain $\hat{a}_{Q_\star}=0.1449(63)$, where the error is for $\Delta\alpha_s^{\overline{\rm MS}}(M_Z)=\pm0.0011$. We then obtain the scheme-and-scale independent PMC prediction on $\delta^{(0)}_4(M_\tau^2,C)$,
\begin{eqnarray}
\delta^{(0)}_4(M_\tau^2,C)|_{\rm PMC} = 0.2035 \pm0.0123 \pm0.0030.
\label{eq:Rtau-PMC}
\end{eqnarray}
where the first error is for $\Delta\alpha_s^{\overline{\rm MS}}(M_Z)=\pm0.0011$ and the second error is an estimate of the ``unknown'' high-order contribution.

\begin{figure}[htb]
\includegraphics[width=0.45\textwidth]{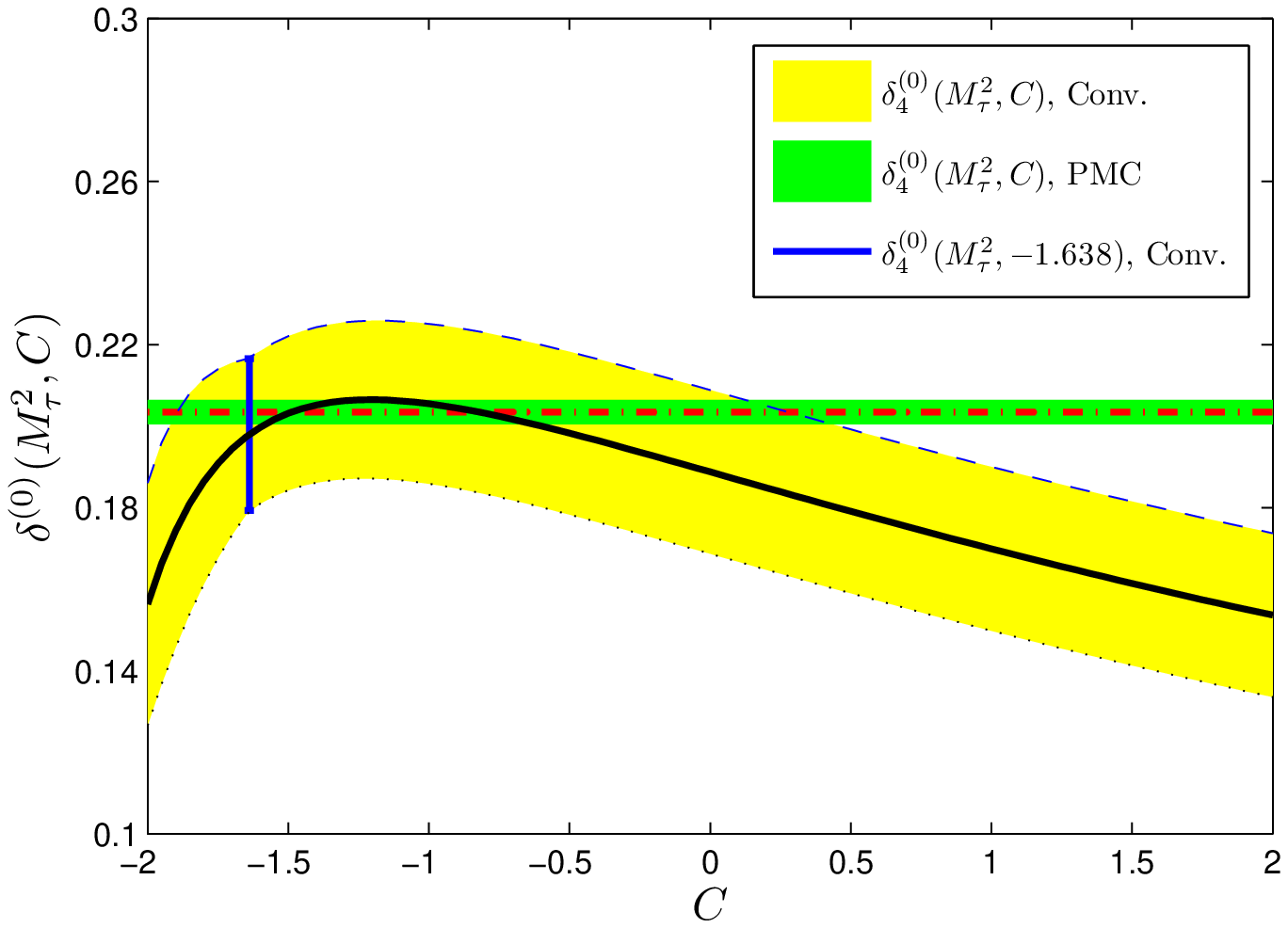}
\includegraphics[width=0.45\textwidth]{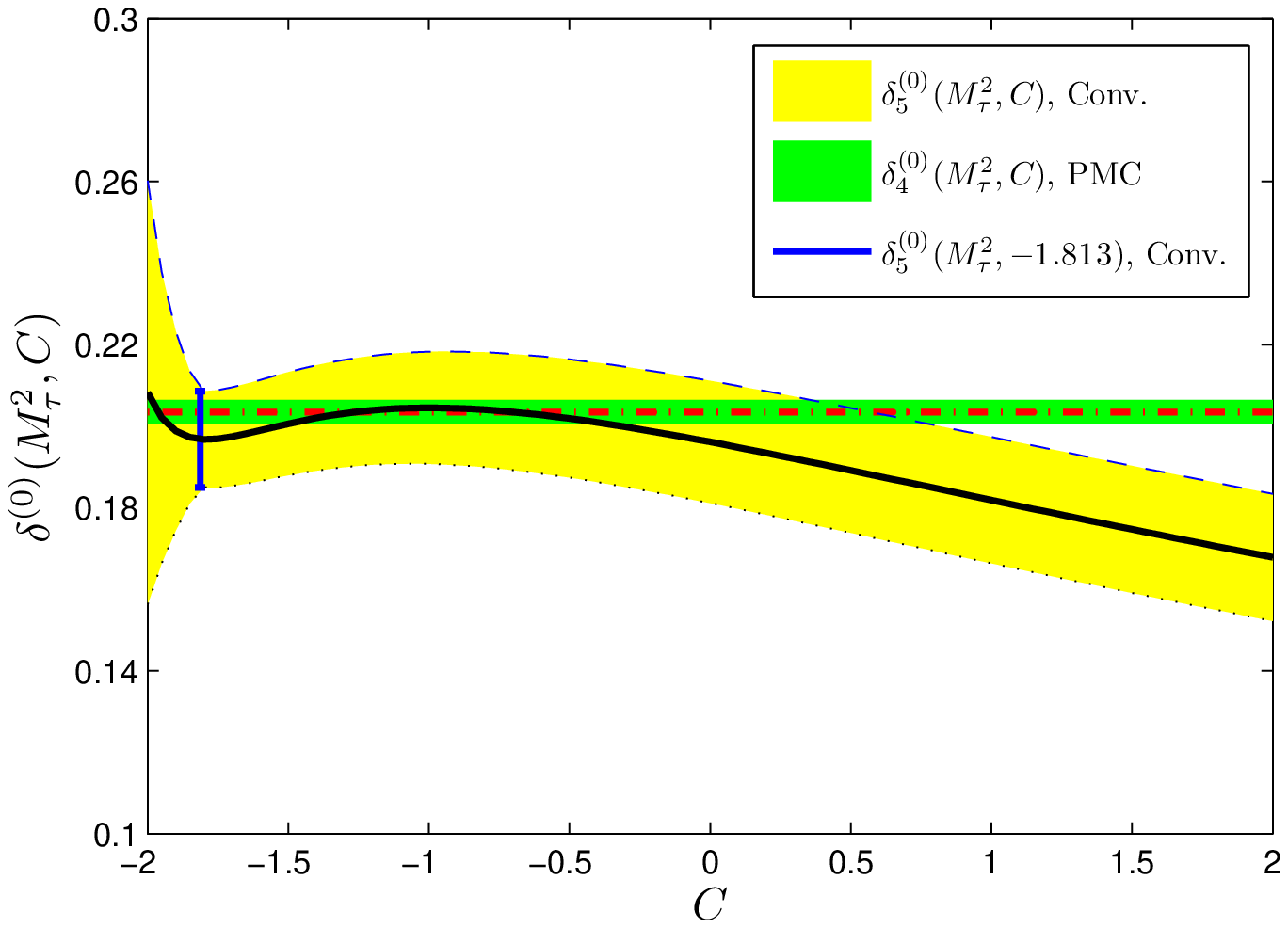}
\caption{$\delta^{(0)}(M_\tau^2,C)$ for $\tau $ decay as a function of the parameter $C$. The solid line is the prediction using conventional scale setting, the lighter-shaded band is the uncertainty for a four-loop prediction $\Delta = \pm|\hat{c}_4(\mu/M_\tau) \hat{a}_{\mu}^4|_{\rm MAX}$ (Upper) and for an approximate five-loop prediction $\Delta = \pm|\hat{c}_5(\mu/M_\tau) \hat{a}_{\mu}^5|_{\rm MAX}$ (Lower), where MAX is the maximum value for $\mu\in[M_\tau,4M_{\tau}]$. When $C=-1.638$ (Upper) and $C=-1.183$ (Lower), the error bar as shown by a vertical solid line is the minimum. The dash-dot line represents the four-loop PMC prediction, and the darker shaded band is for $\Delta = \pm |r_{4,0}\hat{a}^4_{Q_\star}|$ . }
\label{fig:Rtau4l2Dimproved}
\end{figure}

\begin{table*}[htb]
\centering
\caption{The value of each loop-term, LO, NLO, N$^2$LO, or N$^3$LO, for the four-loop prediction $\delta^{(0)}_4(M_\tau^2)$ using conventional (Conv.) and PMC scale-settings, respectively. $\mu=Q=M_\tau$. The results for the $\overline{\rm MS}$-scheme and the optimal $C$-scheme with $C=-1.638$ are presented. The PMC prediction is unchanged for any $C$-scheme. }
\begin{tabular}{ c c c c c c}
\hline
~~~~   & ~LO~ & ~NLO~ & ~N$^2$LO~ & ~N$^3$LO~ & ~Total~  \\ \hline
~Conv., $\overline{\rm MS}$-scheme~  & 0.1006 & 0.0526 & 0.0268 & 0.0130 & 0.1930  \\
~Conv., optimal $C$-scheme~ & 0.1873 & 0.0532 & $-0.0240$ & $-0.0186$ & 0.1979   \\
\hline
~PMC, any $C$-scheme~  & 0.1449 & 0.0451 & 0.0105 & 0.0030 & 0.2035  \\
\hline
\end{tabular}
\label{tab:RtauOrderNew}
\end{table*}

To compare the scheme dependence before and after applying the PMC, we present the predictions on $\delta^{(0)}(M_\tau^2,C)$ for $\mu=M_\tau$ in Fig.~\ref{fig:Rtau4l2Dimproved}. At four-loop level, Fig.~\ref{fig:Rtau4l2Dimproved} shows that the error band shall first increases and then decreases with increasing $C$; the optimal scheme is obtained for $C=-1.638$. By using the approximate five-loop term $r_5\simeq 283$, we give the results for the approximate five-loop prediction. Fig.~\ref{fig:Rtau4l2Dimproved} shows a smaller error bar is achieved with a five-loop term, $\bar{D}^{\rm ns}_5(M_\tau^2,C)|_{\rm Conv.}$ oscillates with the increment of $C$, and the optimal scheme is slightly shifted to $C=-1.813$. Similar to the case of $\bar{D}^{\rm ns}(M_\tau^2,C)|_{\rm Conv.}$, it is observed that the predicted $\delta^{(0)}(M_\tau^2,C)|_{\rm Conv.}$ with the optimal choice of $C$ are also consistent with the PMC predictions within errors. The flat dash-dot line in Fig.~\ref{fig:Rtau4l2Dimproved} shows the conventional scheme dependence can be eliminated by applying the PMC. The unknown high-order contribution could be quite small for the present four-loop PMC prediction.

We present the value of each loop-term, LO, NLO, N$^2$LO, or N$^3$LO, for the four-loop prediction $\delta^{(0)}_4$ using conventional and PMC scale-settings in Table~\ref{tab:RtauOrderNew}. The pQCD convergence for the conventional $\overline{\rm MS}$-scheme is moderate. The pQCD convergence for the optimal $C$-scheme ($C=-1.638$) does not suffer from the usual $\alpha_s$-suppression, $|\delta^{(0),{\rm LO}}_4| \gg |\delta^{(0),{\rm NLO}}_4| \sim |\delta^{(0),{\rm N^{2}LO}}_4| \sim |\delta^{(0),{\rm N^{3}LO}}_4|$. On the other hand, by applying the PMC, a better pQCD convergence is achieved due to the elimination of divergent renormalon-like terms.

\section{Summary}
\label{sec:summary}

We have shown that the scheme-and-scale ambiguities introduced by conventional scale-setting are unnecessary by combining the single-scale PMC procedure with the newly suggested scheme-independent $C$-scheme coupling. We have demonstrated that using the $C$-scheme, together with the PMC-s scale-setting, leads to perturbative QCD predictions which are independent of the initial renormalization scale and the choice of the renormalization scheme at all orders.

The pQCD predictions based on the PMC satisfy the standard RGI and all the self-consistency conditions of the renormalization group. Because the divergent renormalon terms are eliminated, the convergence of the PMC series depends on conformal coefficients, which are generally more convergent than conventional pQCD series. The resulting PMC-s approach not only makes the implementation and automation of the PMC simpler and more transparent, but it also achieves precise scheme-and-scale independent predictions simultaneously.

This method for eliminating the scale and scheme ambiguities relies heavily on how well we know the precise value and analytic properties of the strong coupling. In conventional RGE, the scale running is entangled with the scheme parameters, which can only be solved perturbatively or numerically. In contrast, the scheme-and-scale running behavior of a $C$-scheme coupling $\hat \alpha_s$ can be exactly separated; it satisfies a RGE free of scheme-dependent $\{\beta_{i}\}$-terms. The resulting scheme-independent RGE for the $C$-scheme coupling $\hat \alpha_s$ provides scheme-independent predictions.

As we have shown, one can utilize a novel {\it $C$-scheme} coupling whose scheme-and-scale running behaviors are governed by a single scheme-independent RGE. The value of the parameter $C$ can be chosen to match any conventional renormalization scheme. By using the $C$-scheme coupling instead of conventional coupling, we have demonstrated that the $C$-dependence of the PMC predictions can be eliminated up to any fixed order; since the value of $C$ is arbitrary, it means the PMC prediction is independent of any renormalization scheme. Two four-loop PMC examples confirm those observations. Thus combining the $C$-scheme coupling with the PMC-s approach, the resulting predictions become completely independent of the choice of the renormalization scheme and the initial renormalization scale, satisfying all of the conditions of RGI. The PMC procedure thus systematically eliminates the scheme-and-scale ambiguities of pQCD predictions, greatly improving the precision of tests of Standard Model and the sensitivity of collider experiments to new physics.  \\

\noindent{\bf Acknowledgments: } This work was supported in part by the National Natural Science Foundation of China under Grant No.11625520, and the Department of Energy Contract No.DE-AC02-76SF00515. SLAC-PUB-17232.


\begin{thebibliography}{1}

\bibitem{Beneke:1998ui}
  M.~Beneke,
  Phys.\ Rept.\  {\bf 317}, 1 (1999).

\bibitem{Gardi:2001wg}
  E.~Gardi and G.~Grunberg,
  Phys.\ Lett.\ B {\bf 517}, 215 (2001).

\bibitem{Callan:1970yg}
  C.~G.~Callan, Jr.,
  Phys.\ Rev.\ D {\bf 2}, 1541 (1970).

\bibitem{Symanzik:1970rt}
  K.~Symanzik,
  Commun.\ Math.\ Phys.\  {\bf 18}, 227 (1970).

\bibitem{Stueckelberg:1953dz}
  A.~Petermann,
  Helv.\ Phys.\ Acta {\bf 26}, 499 (1953).

\bibitem{peter2}
	N.~N.~Bogoliubov and D.~V.~Shirkov,
   Dokl. Akad. Nauk SSSR {\bf 103}, 391 (1955).

\bibitem{Peterman:1978tb}
  A.~Peterman,
  Phys.\ Rept.\  {\bf 53}, 157 (1979).

\bibitem{Wu:2013ei}
  X.~G.~Wu, S.~J.~Brodsky and M.~Mojaza,
  Prog.\ Part.\ Nucl.\ Phys.\  {\bf 72}, 44 (2013).

\bibitem{Brodsky:2011ig}
  S.~J.~Brodsky and L.~Di Giustino,
  Phys.\ Rev.\ D {\bf 86}, 085026 (2012).

\bibitem{Brodsky:2011ta}
  S.~J.~Brodsky and X.~G.~Wu,
  Phys.\ Rev.\ D {\bf 85}, 034038 (2012).

\bibitem{Brodsky:2012sz}
  S.~J.~Brodsky and X.~G.~Wu,
  Phys.\ Rev.\ D {\bf 86}, 014021 (2012).

\bibitem{Brodsky:2012rj}
  S.~J.~Brodsky and X.~G.~Wu,
  Phys.\ Rev.\ Lett.\  {\bf 109}, 042002 (2012).

\bibitem{Mojaza:2012mf}
  M.~Mojaza, S.~J.~Brodsky and X.~G.~Wu,
  Phys.\ Rev.\ Lett.\  {\bf 110}, 192001 (2013).

\bibitem{Brodsky:2013vpa}
  S.~J.~Brodsky, M.~Mojaza and X.~G.~Wu,
  Phys.\ Rev.\ D {\bf 89}, 014027 (2014).

\bibitem{Brodsky:2012ms}
  S.~J.~Brodsky and X.~G.~Wu,
  Phys.\ Rev.\ D {\bf 86}, 054018 (2012).

\bibitem{Brodsky:1982gc}
  S.~J.~Brodsky, G.~P.~Lepage and P.~B.~Mackenzie,
  Phys.\ Rev.\ D {\bf 28}, 228 (1983).

\bibitem{GellMann:1954fq}
  M.~Gell-Mann and F.~E.~Low,
  Phys.\ Rev.\  {\bf 95}, 1300 (1954).

\bibitem{Brodsky:1997jk}
  S.~J.~Brodsky and P.~Huet,
  Phys.\ Lett.\ B {\bf 417}, 145 (1998).

\bibitem{Shen:2017pdu}
  J.~M.~Shen, X.~G.~Wu, B.~L.~Du and S.~J.~Brodsky,
  Phys.\ Rev.\ D {\bf 95}, 094006 (2017).

\bibitem{Brodsky:1994eh}
  S.~J.~Brodsky and H.~J.~Lu,
  Phys.\ Rev.\ D {\bf 51}, 3652 (1995).

\bibitem{Shen:2016dnq}
  J.~M.~Shen, X.~G.~Wu, Y.~Ma and S.~J.~Brodsky,
  Phys.\ Lett.\ B {\bf 770}, 494 (2017).

\bibitem{Boito:2016pwf}
  D.~Boito, M.~Jamin and R.~Miravitllas,
  Phys.\ Rev.\ Lett.\  {\bf 117}, 152001 (2016).

\bibitem{Gross:1973id}
  D.~J.~Gross and F.~Wilczek,
  Phys.\ Rev.\ Lett.\  {\bf 30}, 1343 (1973).

\bibitem{Politzer:1973fx}
  H.~D.~Politzer,
  Phys.\ Rev.\ Lett.\  {\bf 30}, 1346 (1973).

\bibitem{Caswell:1974gg}
  W.~E.~Caswell,
  Phys.\ Rev.\ Lett.\  {\bf 33}, 244 (1974).

\bibitem{Tarasov:1980au}
  O.~V.~Tarasov, A.~A.~Vladimirov and A.~Y.~Zharkov,
  Phys.\ Lett.\  B {\bf 93}, 429 (1980).

\bibitem{Larin:1993tp}
  S.~A.~Larin and J.~A.~M.~Vermaseren,
  Phys.\ Lett.\ B {\bf 303}, 334 (1993).

\bibitem{vanRitbergen:1997va}
  T.~van Ritbergen, J.~A.~M.~Vermaseren and S.~A.~Larin,
  Phys.\ Lett.\ B {\bf 400}, 379 (1997).

\bibitem{Chetyrkin:2004mf}
  K.~G.~Chetyrkin,
  Nucl.\ Phys.\ B {\bf 710}, 499 (2005).

\bibitem{Czakon:2004bu}
  M.~Czakon,
  Nucl.\ Phys.\ B {\bf 710}, 485 (2005).

\bibitem{Baikov:2016tgj}
  P.~A.~Baikov, K.~G.~Chetyrkin and J.~H.~Kuhn,
  Phys.\ Rev.\ Lett.\  {\bf 118}, 082002 (2017).

\bibitem{Corless:1996zz}
  R.~M.~Corless, G.~H.~Gonnet, D.~E.~G.~Hare, D.~J.~Jeffrey and D.~E.~Knuth,
  Adv.\ Comput.\ Math.\  {\bf 5}, 329 (1996).

\bibitem{Bi:2015wea}
  H.~Y.~Bi, X.~G.~Wu, Y.~Ma, H.~H.~Ma, S.~J.~Brodsky and M.~Mojaza,
  Phys.\ Lett.\ B {\bf 748}, 13 (2015).

\bibitem{Jamin:2016ihy}
  M.~Jamin and R.~Miravitllas,
  JHEP {\bf 1610}, 059 (2016).

\bibitem{Stevenson:1980du}
  P.~M.~Stevenson,
  Phys.\ Lett.\ B {\bf 100}, 61 (1981).

\bibitem{Stevenson:1981vj}
  P.~M.~Stevenson,
  Phys.\ Rev.\ D {\bf 23}, 2916 (1981).

\bibitem{Stevenson:1982qw}
  P.~M.~Stevenson,
  Nucl.\ Phys.\ B {\bf 231}, 65 (1984).

\bibitem{Ma:2014oba}
  Y.~Ma, X.~G.~Wu, H.~H.~Ma and H.~Y.~Han,
  Phys.\ Rev.\ D {\bf 91}, no. 3, 034006 (2015).

\bibitem{Wu:2014iba}
  X.~G.~Wu, Y.~Ma, S.~Q.~Wang, H.~B.~Fu, H.~H.~Ma, S.~J.~Brodsky and M.~Mojaza,
  Rept.\ Prog.\ Phys.\  {\bf 78}, 126201 (2015).

\bibitem{Olive:2016xmw}
  C.~Patrignani {\it et al.} [Particle Data Group],
  Chin.\ Phys.\ C {\bf 40}, 100001 (2016).

\bibitem{Adler:1974gd}
  S.~L.~Adler,
  Phys.\ Rev.\ D {\bf 10}, 3714 (1974).

\bibitem{Baikov:2012zm}
  P.~A.~Baikov, K.~G.~Chetyrkin, J.~H.~Kuhn and J.~Rittinger,
  JHEP {\bf 1207}, 017 (2012).

\bibitem{Chetyrkin:1996ia}
  K.~G.~Chetyrkin, J.~H.~Kuhn and A.~Kwiatkowski,
  Phys.\ Rept.\  {\bf 277}, 189 (1996).

\bibitem{Baikov:2010je}
  P.~A.~Baikov, K.~G.~Chetyrkin and J.~H.~Kuhn,
  Phys.\ Rev.\ Lett.\  {\bf 104}, 132004 (2010).

\bibitem{Beneke:2008ad}
  M.~Beneke and M.~Jamin,
  JHEP {\bf 0809}, 044 (2008).

\bibitem{Baikov:2008jh}
  P.~A.~Baikov, K.~G.~Chetyrkin and J.~H.~Kuhn,
  Phys.\ Rev.\ Lett.\  {\bf 101}, 012002 (2008).

\end{thebibliography}
\end{document}